\documentclass[a4paper]{llncs}
\usepackage{algorithm}
\usepackage{algorithmic}
\usepackage[labelformat=simple]{subfig}
\usepackage{setspace}
\usepackage{amsmath}
\usepackage{amssymb}
\usepackage{multirow}
\usepackage{fancyvrb}
\usepackage{xcolor}
\usepackage{xspace}
\usepackage{rotating}
\usepackage{wrapfig}
\usepackage{cite}
\usepackage{pgfplots}
\usetikzlibrary{patterns,trees,shapes,positioning}
\pgfplotsset{width=15cm,compat=1.5}
\usepackage{url}
\urldef{\mailsa}\path|sbjoshi@cse.iitk.ac.in|
\urldef{\mailsb}\path|akashl@microsoft.com|

\newcommand{\figref}[1]{Fig.~\ref{Fi:#1}}
\newcommand{\figrefs}[2]{Figs.~\ref{Fi:#1} and~\ref{Fi:#2}}

\renewcommand{\eqref}[1]{Eqn.~(\ref{Eq:#1})}

\newcommand{\tableref}[1]{Tab.~\ref{Ta:#1}}

\newcommand{\sectref}[1]{Section~\ref{Se:#1}}
\newcommand{\sectrefs}[2]{Sections~\ref{Se:#1} and~\ref{Se:#2}}

\newcommand{\algref}[1]{Alg.~\ref{Alg:#1}}
\newcommand{\algrefs}[2]{Algs.~\ref{Alg:#1} and~\ref{Alg:#2}}
\newcommand{\algrefsp}[3]{Algs.~\ref{Alg:#1}, \ref{Alg:#2}, and~\ref{Alg:#3}}

\newcommand{\inangleb}[1]{ \ensuremath{\langle #1 \rangle}}

\newcommand{\inroundb}[1]{\ensuremath{ \left( #1 \right)}}
\newcommand{\inparan}[1] {\ensuremath{ \{ #1 \}} }
\newcommand{\SC}[1]{\textsc{#1}}
\newcommand{\SSZ}[1]{\scriptsize{#1}}
\newcommand{\IT}[1]{\textit{#1}}
\newcommand{\TT}[1]{\texttt{#1}}
\newcommand{\MB}[1]{\ensuremath{\mathbf{#1}}}
\newcommand{\BF}[1]{\textbf{#1}}
\newcommand{\Omit}[1]{}

\newcommand{\assume}{\mathbf{assume}}
\newcommand{\yield}{\mathbf{yield}}
\newcommand{\axiom}{\mathbf{axiom}}
\newcommand{\satomic}{\mathbf{satomic}}
\newcommand{\watomic}{\mathbf{watomic}}
\newcommand{\async}{\mathbf{async}}
\newcommand{\join}{\mathbf{join}}

\newcommand{\tool}{\textsc{AtomicInf}\xspace}

\newcommand{\CSG}{\textit{CSG}}
\newcommand{\tup}[1]{\langle #1 \rangle}

\newcommand{\eqdef}{\buildrel \mbox{\tiny\rm def} \over =}

\DeclareCaptionSubType*{figure}

\makeatletter
\newbox\sf@box
\newenvironment{SubFloat}[2][]%
{\def\sf@one{#1}%
\def\sf@two{#2}%
\setbox\sf@box\hbox
\bgroup}%
{ \egroup
\ifx\@empty\sf@two\@empty\relax
\def\sf@two{\@empty}
\fi
\ifx\@empty\sf@one\@empty\relax
\subfloat[\sf@two]{\box\sf@box}%
\else
\subfloat[\sf@one][\sf@two]{\box\sf@box}%
\fi}
\makeatother

\begin{document}

\mainmatter  

\title{Automatically finding atomic regions for fixing bugs in Concurrent programs}


%
%
\author{Saurabh Joshi\inst{1}
\and Akash Lal\inst{2}}%
%
\authorrunning{ }

\institute{Department of CSE, IIT Kanpur, India\\
\and Microsoft Research, India\\
}

%
%

\maketitle
\begin{abstract}
This paper presents a technique for automatically constructing a fix for
buggy concurrent programs: given a concurrent program that does not
satisfy user-provided assertions, we infer atomic blocks that fix the
program. 
An atomic block protects a piece of code and ensures that it
runs without interruption from other threads. 
Our technique uses a verification tool as a subroutine to find the
smallest atomic regions that remove all bugs in a given
program. Keeping the atomic regions small allows for maximum
concurrency. We have implemented our approach in a tool called \tool. 
A user of \tool can  choose between strong and weak atomicity semantics for the
inferred fix. While the former is simpler to find, the latter 
provides more information about the bugs that got fixed.

We ran \tool on several benchmarks and came up with the
smallest and the most precise atomic regions in all of them.
We implemented an earlier technique to our setting and observed that \tool is 1.7 times
faster 
on an average as compared to an earlier approach.

\end{abstract}


\section{Introduction}

An important part of the debugging process is to come up with a repair that fixes the bug under
investigation. After a candidate repair is formulated, not only must one reason that the fix removes the bug, 
but also that it does not introduce new bugs in the program. Thus, evaluating a repair requires
understanding of the program as a whole, not just the executions that reveal the current bug.
Consequently, any automation in the process of formulating and evaluating a fix would be welcome.

At first, this debugging problem seems to be a good match for verification tools that are prepared to 
reason over many (or all) program behaviors. However, the process of formulating a fix can be
difficult to automate. For instance, any program with an assertion \TT{assert(!error)} can be ``fixed'' by
inserting the statement \TT{error=false} right before the assert. However, such repairs are clearly of no 
practical use. 

In order to get across this challenge of meaningless repairs, we focus on a restricted problem
of automated program repair. In particular, we focus on fixing concurrent programs by introducing extra
synchronization that restricts the set of interleavings possible in the program. \emph{Given a specification,
we infer a fix that removes \IT{all} of the bad interleavings of the program while \IT{minimizing} the set of
good interleavings that get removed by the fix.} This problem definition has several advantages.
First, a fix is not allowed to introduce new behaviors, i.e., any execution of the fixed program is also
a valid execution of the original program. This rules out, for instance, the trivial repair mentioned
previously. Second, we remove all bad interleavings, which implies that the resulting program will
satisfy the specification. Third, by minimizing the set of good interleavings removed, we allow maximum
concurrency in the program and avoid significantly degrading the performance and responsiveness of the program.
Fourth, the specification is supplied by the user, allowing one to target the repair towards certain (user-defined)
properties.

We restrict the space of interleavings by introducing extra synchronization in the program
in the form of \IT{atomic blocks}. Atomic blocks are a convenient way of expressing synchronization.
Previous work shows that programs that use atomic blocks are easier to understand than ones
that use locks \cite{rossbach10ppopp}. An atomic block is used to enclose a piece of code that restricts how 
that code interacts with concurrently executing threads. The exact semantics depends on the type
of atomic block used. A \IT{strong} atomic block ensures that the enclosed code executes in complete isolation of the
rest of the program. A \IT{weak} atomic block ensures that the enclosed code 
executes in isolation of other weak atomic sections. We allow a user to pick which kind of atomic blocks
to use for the fix. A fix using strong atomic blocks is easier to find, but a fix using weak atomic blocks
usually reveals more information about the bug getting fixed. Furthermore, it is easy to realize weak 
atomic blocks using locks \cite{emmi07popl}.

Our approach works as follows. We accept, as input, a program with assertions. We assume that the
program specification is fully captured by the asserts. (Any safety property can be captured
using assertions.) \emph{Furthermore, we assume that all executions of the program in which threads
do not interleave (i.e., the threads execute sequentially, one after the other) are
correct.} 
This is an important assumption because otherwise, no set of atomic blocks could repair the program.
Next, we use any off-the-shelf verification tool to iteratively reveal more and more buggy traces in the program
until we converge on a fix. Because queries to the verification tool can be expensive, we minimize the
number of buggy traces required. 

When a user selects strong atomicity, we guarantee that the reported fix is the smallest in terms of the number of
program points protected by the atomic block. However, the search for the smallest weak atomic blocks turns
out to be too expensive. Thus, when a user selects weak atomicity, we employ a crucial optimization. We first
find the smallest fix $F$ under strong atomicity and then restrict the search for weak atomic
blocks to those which are
 supersets of $F$. While this implies that the fix may not be the smallest, we still
guarantee that it is a minimal extension  of $F$. Furthermore, our experiments reveal that this optimization does not 
compromise the quality of the fix.

The key contributions of this paper are as follows:
\begin{itemize}
\item We give an efficient approach for finding a smallest fix $F$ under strong atomicity as well
as a minimal extension of  $F$ under weak atomicity.
\item Our approach is completely driven by user-supplied properties, as opposed to
previous work
that relies on symptoms such as data races and atomicity violations to root-cause a bug \cite{flanagan04atomizer,eraser}. 
Hence, our technique does not get distracted by benign data races.
\item Our experiments show that we are able to find the best fix on a variety of benchmarks.
\end{itemize}

The rest of the paper is organized as follows. Section~\ref{Se:Overview} gives an overview of our technique.  Section~\ref{Se:Related} describes the
related work. 
Sections~\ref{Se:Technical} to ~\ref{Se:Weak} describe our techniques in detail with algorithms and
proofs. Section~\ref{Se:Experiments} mentions experimental results.

\section{Overview}
\label{Se:Overview}

This section gives an overview of \tool. It accepts as input a concurrent program with assertions.
We use the term \textit{bug} to refer to an execution that ends in an assertion violation.

\begin{sloppypar}
We consider two semantics for atomic blocks. A \IT{strong atomic block} \MB{satomic}($S$) guards a
region of code $S$ and ensures that it runs in complete isolation with respect to other threads.
Intuitively, this means that context switching is disabled while $S$ is executing. 
A \IT{weak atomic block} \MB{watomic}($S$) ensures that $S$ runs in isolation with respect to other
weak atomic blocks. One simple (semantics preserving) implementation of weak atomic blocks is to
use a single global lock $l$ and replace each block \MB{watomic}($S$) with (acquire($l)$; $S$; release($l$)).
In this sense, it is much easier to realize weak atomic blocks in a language runtime and people
have proposed efficient implementations for them \cite{emmi07popl}.
\end{sloppypar}

Consider the \texttt{banking} program shown in \figref{banking}. The procedure
\texttt{transfer} transfers a given amount from one account
to another account. The procedure \texttt{seize} sets the balance in a given
account to zero. The thread \texttt{thread1}
attempts to transfer 200 units from \texttt{acc1} to \texttt{acc2};
\texttt{thread2} tries to set the balance in \texttt{acc1} to zero;
and \texttt{thread3} tries to transfer 100 units from
\texttt{acc2} to \texttt{acc1}. The program is buggy: for example, 
immediately after \texttt{thread1} checks whether sufficient
amount is available in \texttt{acc1} in \texttt{transfer},
\texttt{thread3} starts and runs to completion, then
\texttt{thread2} also runs to completion, setting the balance in
\texttt{acc1} to \texttt{0}. Next, when \texttt{thread1} finishes, the
amount in \texttt{acc1} is negative. The appropriate fixes for the program
are the following: (1) enclose the body of \texttt{transfer} in a strong
atomic block; or (2) enclose the bodies of both \texttt{transfer} and 
\texttt{seize} in weak atomic blocks. Note that the latter solution is more
informative because it points out the conflicting concurrent accesses. 
\tool can find these fixes automatically.

\begin{figure}[t]
\begin{tabular}{p{5cm}p{1ex}}
\begin{scriptsize}
\begin{Verbatim}[numbers=left]
struct Account {
    int amount;
} acc1, acc2, acc3;

void seize(Account *acc) { 
  acc->amount = 0; 
}

void thread1() {
  transfer(&acc1,&acc2,200);
}
void thread2() {
  seize(&acc1);
}
void thread3() {
  transfer(&acc2,&acc1,100);
}
\end{Verbatim}
\end{scriptsize}
& 
\begin{scriptsize}
\begin{Verbatim}[numbers=left,firstnumber=18]
int transfer(Account* src,Account* dst,int amount) {
      if(src->amount>=amount) {
        int temp = src->amount;
        temp = temp - amount;
        src->amount = temp;
        temp = dst->amount;
        temp = temp + amount;
        dst->amount = temp;
        return 1;
      }
      return 0; 
}
void main() {
  acc1.amount = acc2.amount = acc3.amount = 200;
  t1 = async thread1();
  t2 = async thread2();
  t3 = async thread3();
  join(t1); join(t2); join(t3);
  assert(acc1.amount == 0 || acc1.amount == 100);
}
\end{Verbatim}
\end{scriptsize}
\end{tabular}
\caption{The Banking example. The \texttt{async} keyword spawns a thread and returns a handle for the thread. The
\texttt{join} command waits for the thread to terminate.}
\label{Fi:banking}
\end{figure}


Let \SC{VerifTool} refer to any verification tool that gives us the ability of controlling the places where context 
switches are allowed. Let $P$ be the buggy concurrent program.
We feed $P$ to \SC{VerifTool} and get a trace $t$ that witnesses a buggy execution of $P$. 
We examine the set of program locations where the trace took context switches. Lets say
this set is $\{l_1, l_2, l_3 \}$. Then we add the constraint ``$\neg(l_1 \wedge l_2 \wedge l_3)$''
to \SC{VerifTool}, i.e., we disallow it from considering traces that take context switches at all of these locations. 
This constraint rules out $t$ from appearing again.
Next, we feed $P$ to \SC{VerifTool} under this constraint. We repeat this process until we get no more 
buggy traces. Suppose that $L_i$ was the set of context switch locations of the $i^\text{th}$
buggy trace. Then a solution, for strong atomicity, is to select at least one location from each
$L_i$, and enclose all such locations in strong atomic blocks. This is sound because the fix disables
all buggy traces of $P$. This technique is inspired
from previous work \cite{yahav10popl}. As we describe later, this technique is very inefficient
because the number of queries made to \SC{VerifTool} can be proportional to the size of the program, irrespective
of the size of the solution. For example, suppose the fix requires protecting just one location $l$.
Then there would still be lots of buggy traces that take a context switch at $l$ but also take
(possibly redundant) context switches at other locations. The above iterative process will enumerate
all such traces.

We propose an algorithm where the number of queries to \SC{VerifTool}
remains roughly proportional to the size of the solution.
In order to accomplish this, we use the same iterative scheme as above,
but after the $i^\text{th}$ iteration, we take the $i$ buggy traces and construct
a ``proposed'' solution by selecting at least one context-switch location from each 
of the $i$ traces. Next, we ask \SC{VerifTool} if this solution fixes the program. If so, we're done.
Otherwise, we get a buggy trace and we repeat the process. This way, redundant context
switches get ruled out very quickly. For example, if the fix is one location $l$, then
after getting two traces that take context switches $\{l, l_1\}$ and $\{l, l_2\}$, the 
proposed solution of $\{ l \}$ will be the correct fix, without requiring enumeration of
other buggy traces. 

For weak atomicity, we follow a similar iterative process, but mine the
error traces for more information. In particular, we look for pairs of locations $(l_1, l_2)$
such that a buggy trace takes a context switch at $l_1$ and passes through $l_2$. We can
rule out this error trace by placing both locations inside a weak atomic block. 
To reduce the search space, we first find a solution under strong atomicity and then use it
as a starting point for finding a solution under weak atomicity. In many cases, these solutions are very similar,
thus, reducing the number of iterations required for weak atomicity. For the \texttt{banking}
example, once the strong atomicity fix is found, extending it to a weak atomicity fix only
requires enclosing the \texttt{seize} procedure in a weak atomic block.

We now illustrate the property-guided nature of \tool.
For the program of \figref{banking}, 
the fix reported by \tool 
allowed a context switch at line
\ref{csallowed}, as shown in \figref{toolfix}. On closer inspection, this is a valid solution;
it says that operations of debiting \TT{amount} from \TT{src}, and crediting to \TT{dst} need
to be individually atomic, but it is fine for other operations to execute between them.



As a further test, we changed the implementation of the second thread to what is shown in 
\figref{bankingcorpus}. It checks that the corpus of money in the two accounts remains constant.
Because this is done in a thread, the assertion can fire any time during the program's execution.
In this case, \tool proposes that the entire body of \TT{transfer} needs to be inside a single
strong atomic block; it is no longer safe to interleave operations between the debit and credit
of \texttt{transfer}.

\begin{figure}[t]
\begin{SubFloat}{\label{Fi:toolfix}}
\begin{minipage}{0.5\linewidth}
\begin{scriptsize}
\begin{Verbatim}[numbers=left,commandchars=*\[\]]
int transfer(Account* src,Account* dst,int amount)
{
  satomic
   {
      if(src.amount>=amount) *label[atom1start]
      {
        int temp = src.amount;
        temp = temp - amount;
        src.amount = temp; *label[atom1end]
   --> *label[csallowed]     
        temp = dst.amount; *label[atom2start]
        temp = temp + amount; *label[atom2end]
        dst.amount = temp
        return 1;
      }
   }
      return 0;
}
\end{Verbatim}
\end{scriptsize}
\end{minipage}
\end{SubFloat}
\hspace{0.7cm}
\begin{SubFloat}{\label{Fi:bankingcorpus}}
\begin{minipage}{0.45\linewidth}

\begin{scriptsize}
\begin{Verbatim}[numbers=left,commandchars=&\[\]]
void thread2()
{
   int temp=0;
   satomic{
	   temp = acc1.account;
           temp += acc2.account;
           assert(temp == 400);
   }
}

\end{Verbatim}
\end{scriptsize}
\vspace{0.695in}

\end{minipage}
\end{SubFloat}
\caption{\ref{Fi:toolfix} The strong atomicity fix found by \tool for the program in \figref{banking}.
\ref{Fi:bankingcorpus} Asserting that the corpus of the bank must remain constant at all times.}
\end{figure}

\subsection{Related Work}\label{Se:Related}

Automatic repair of programs has been studied earlier, both for sequential programs
\cite{jobstmann2005cav,griesmayer06cav,chandra2011icse,malik2011icst}
as well as for concurrent programs
\cite{cern2011cav,yahav10popl,wypiwyg,atomtracker,jin2011pldi}.
Previous work on sequential programs has focused on formulating program repair
as a two-player game, where one of the players tries to make
sure that the program doesn't fail. A winning strategy for this player is the
repair. One limits the vocabulary of the player (to, for instance, memory-less
players) in order to reduce the search space and come up with a reasonable fix.
This work is orthogonal to ours because we do not try to repair sequential executions
of a program.


For concurrent programs, a majority of the work uses dynamic analysis
to repair bugs like atomicity violations  
\cite{jin2011pldi,recon2011pldi,burnim2011asplos}. For instance, \SC{Recon}
\cite{recon2011pldi} uses test runs to locate bugs and then uses statistical
analysis over these runs to infer a fix.
Being dynamic in nature allows these techniques to scale, but the quality
of the solution is dependant on the coverage of the test runs.
On the other hand, our approach uses static analysis and is capable of providing
soundness guarantees for the fix. Moreover, our notion of
a bug is an assertion failure, not notions like data-races and
atomicity violations. Thus, our approach does not get distracted
by benign (and intended) data-races and atomicity violations.

Some techniques require user annotations to infer necessary
synchronization. For example, the approach described in \cite{vaziri2006popl} 
infers synchronization once a user annotates sets of fields, indicating 
existence of a consistency property within members of each set.

A quantitative approach to synthesize synchronization has been
proposed in \cite{cern2011cav}. This work tries to optimize the
synthesis of synchronization with respect to a performance
model. Though this work provides correctness as well as performance
guarantees about the fix, it only works for finite-state programs,
making its use very limited.


We now discuss two pieces of work that are most similar to ours.
First is \SC{Wypiwyg} (What-You-Prove-Is-What-You-Get) \cite{wypiwyg}, which takes a correct sequential library
and then synthesizes synchronization (in the form of locks) to make sure that the library functions
correctly even in the presence of a concurrent client. Their idea is to take the proof-of-correctness
under a sequential client and then construct synchronization to preserve the same proof even under a
concurrent client. This approach contrasts with ours in the following ways: First, \tool relies on a bug-finding
tool, not necessarily ones that can produce a proof of correctness. Second, \tool guarantees
to find the smallest fix (under atomic sections) irrespective of the underneath verification tool, 
whereas the quality of the solution in \SC{Wypiwyg} depends
completely on the quality of the proof produced---the more modular the proof, the better the synchronization inferred.
We ran \tool on the benchmarks used by \SC{Wypiwyg}. Both approaches inferred the ideal synchronization.
However, it is not possible to compare the running times because \SC{Wypiwyg} used a manually-constructed
proof of correctness for some benchmarks.

The work by Vechev et al. \cite{yahav10popl} is also very similar to
ours. The goal of their work was to exhibit the power of
abstraction-refinement for synthesizing synchronization using strong atomic blocks.
We recast their approach to our setting in \sectref{mhs}, and then show that our 
technique (\sectref{optmhs}) is more efficient. Moreover, their work did not address
inferring synchronization under weak atomicity.

\section {Preliminaries}
\label{Se:Technical}

\newcommand{\CS}{\text{CS}}
\newcommand{\WCS}{\text{WCS}}
\newcommand{\verify}{\textsc{VerifTool}}
\newcommand{\correct}{\textsc{Correct}}
\newcommand{\bug}{\textsc{Bug}}
\newcommand{\lock}{\texttt{lock}}

This section sets up the program syntax used in the rest of the paper and the problem definition.
Because we want to control context switching in the program, we
assume a co-operative model of concurrency where a program is only allowed to take a context switch at a special
$\yield$ instruction. We write programs using C syntax, extended with the following constructs. 

\begin{description}

\item [$\yield$]: The program can context switch only at this statement. 

\item [$\assume(e)$]: If the expression $e$ evaluates to \textit{false} then the program blocks, otherwise
it continues to the next statement.
 
\item [$\axiom(e)$]: This statement is similar to having $\assume(e)$ at all points in the program. We use $\axiom$ to insert global invariants into a program.

\item [$\satomic (stmt)$]: This specifies a strong atomic region. $stmt$ is executed atomically, i.e., no context switches are allowed while executing $stmt$.

\item [$\watomic (stmt)$]: This specifies weak atomic region. $stmt$ is executed in isolation with respect to all
other $\watomic$ blocks. In other words, the execution of $stmt$ can not begin if some other thread is
executing inside a $\watomic$ block.

\item [$\async$ $m()$]: This construct spawns a thread which executes method $m()$. It also returns a handle of the thread created.

\item [$\join (tid)$]: This statement waits for the thread, represented by its handle $tid$, to terminate.
\end{description}

Using the co-operative model of concurrency is not restrictive. Given a multi-threaded program $P$, one can
insert $\yield$ instructions before any instruction that accesses a shared memory location, and also as 
the first instruction of a thread. The resulting program, under co-operative semantics, is equivalent to $P$.
For example, the left side of \figref{transform} shows how the \texttt{transfer} method of \figref{banking} is 
instrumented for co-operative semantics. (For simplicity, we assume that each line of code
executes atomically.) Based on this model, we define the notion of a \IT{minimum fix} as
follows.

\begin{definition}
A \textbf{minimum fix} for a program $P$ is one which encloses the least number of $\yield$
statements under strong or weak atomic blocks.
\end{definition}

\Omit{
Under this model, we define the problem of inferring a fix as one that encloses the least number of
$\yield$ statements under strong or weak atomic blocks.}

 \sectrefs{StrongAtomicity}{Weak}  address problems of finding a fix under strong or weak
atomicity semantics
respectively.

Once we have a fix under the co-operative model, we map the fix back to one in the multi-threaded model.
Let $Y$ be the set of yield instructions that need to be protected by an atomic block, and let $S$
be the set of original program locations where these instructions were inserted.
Next, we say that two statements $stmt_1$ and $stmt_2$ are \textit{connected} if there is
a path from $stmt_1$ to
$stmt_2$ or from $stmt_2$ to $stmt_1$ in the control flow graph of the program, such
that this path does not pass through any program point $p \notin S$. We compute such
maximally connected components within the CFG and output it as the atomic blocks. These
regions may not be lexically scoped. One way to make them lexically scoped is to consider
the set of statements that falls between the dominator and the postdominator of the
maximally connected component found earlier. 
It is a matter of choice whether to
output a maximally connected component as a region or augment it to make it lexically
scoped. 

\subsubsection{Limitations}
Although our algorithms guarantee to find the least number of program points
to protect in a fix, the process of actually reporting atomic blocks may lose this guarantee;
finding lexically-scoped blocks can force us to include other program points in the atomic blocks.
However, this is not a major limitation. \tool also reports the collection of program points and it is usually
easy to manually infer the desired fix from this collection of points.

Another limitation is that, in general, the fix inferred by \tool can only guarantee correctness with respect to safety
properties. \emph{It cannot handle liveness properties.} This limitation shows up when the input program
itself has some synchronization. Then, imposing the fix inferred by \tool can lead to deadlocks. For instance,
if the program uses flag-based synchronization via a loop: \texttt{while(!flag) \{ \}}, (i.e., a thread
waits for some other thread to set \texttt{flag} to true), then disabling context switches within the body
of this loop can cause a deadlock. We circumvent this problem by never including \textbf{yield} 
instructions that are meant for synchronization in our fix. 
This is done partly automatic: \textbf{yield} statements before
synchronization operations such as locking routines, and just after an \textbf{async} are excluded from the fix; 
and partly manual: a user annotates explicit \textbf{yield} points inside shared-memory based 
synchronization operations, which are also excluded from the fix.
We leave a more detailed study for fixing liveness properties as future work.

\section{Strong Atomicity Inference}
\label{Se:StrongAtomicity}

Our first step is to gain control over context switching in the program. We do this by introducing a
fresh Boolean constant for each $\yield$ instruction (except ones excluded because of synchronization---see \sectref{Technical}), 
and then guard the $\yield$s using this
constant as shown in \figref{transform}. Let $\CSG$ be the set of Boolean constants introduced this way.
Forcing a Boolean constant $\textit{cs}_i \in \CSG$ to be $\textit{false}$ will prevent
the context switch
from happening at corresponding $\yield$ point. For example, in \figref{transform} if we want
\texttt{src->amount} to be decremented atomically, we add $\axiom(\textit{cs}_3 == \textit{false})$ to the program.
We also use these Boolean constants to identify the location of a $\mathbf{yield}$ instruction.

\begin{figure}[t]
\begin{tabular}{|p{6cm}|p{6cm}|}
\hline 
Before & After \\ \hline
\vspace{-.5cm}
\begin{scriptsize}
\begin{Verbatim}
int transfer(Account* src,Account* dst,
             int amount) {
   yield;
   if(src->amount >= amount) {
     yield;
     int temp = src->amount;
     temp = temp - amount;
     yield;
     src->amount = temp;
     yield;
     temp = dst->amount;
     temp = temp + amount;
     yield;
     dst->amount = temp;
     return 1;
   }
   return 0;
}
\end{Verbatim}
\end{scriptsize}
 & 
\vspace{-.5cm}
\begin{scriptsize}
\begin{Verbatim}
int transfer(Account* src,Account* dst,
             int amount) {
   if(cs1) { yield; }
   if(src->amount >= amount) {
     if(cs2) { yield; }
     int temp = src->amount;
     temp = temp - amount;
     if(cs3) { yield; }
     src->amount = temp;
     if(cs4) { yield; }
     temp = dst->amount;
     temp = temp + amount;
     if(cs5) { yield; }
     dst->amount = temp;
     return 1;
   }
   return 0;
}
const bool cs1,cs2,cs3,cs4,cs5;
\end{Verbatim}
\end{scriptsize}
\vspace{-.5cm}
\\ \hline
\end{tabular}

\caption{Transforming a program to guard yields}
\label{Fi:transform}
\end{figure}

Given a formula $\phi$ over $\CSG$, let $\tup{P,\phi}$ be the program $P$ extended
with the statement $\axiom(\phi)$. If $S \subseteq \CSG$, then let 
$\textit{disable}(S) = \bigwedge _{cs_i \in S} \neg cs_i$. Our goal is to find
the smallest set $S$ such that $\tup{P, \textit{disable}(S)}$ is a correct program.

For a trace $t$, let $\CS(t) \subseteq \CSG$ be the set of Boolean
constants corresponding to the context switches taken in $t$. Note
that $\CS(t)$ cannot be empty when $t$ is an error trace of a program
without sequential bugs. As previously noted, we assume that the program does not have any
sequential bugs. 
Let \SC{BTraces}($P$) be the (possibly infinite) set of all error
traces of program $P$. Let \SC{CSTraces}($P$) be $\{ \CS(t) ~|~ t \in
\SC{BTraces}(P) \}$. Thus, $\SC{CSTraces} \subseteq (\mathcal{P}(CSG)
\setminus \{ \emptyset \})$, where $\mathcal{P}$ denotes the power set
of a given set. Since $CSG$ is finite, \SC{CSTraces} will be finite as well.

For a program $P$, a valid fix is one that rules out all traces in $\SC{BTraces}(P)$.
To disallow a trace $t$, it is sufficient to disable any one of the context
switches taken by $t$. 
Thus, a fix for $P$ is to disable a set of context switches $S$ such that $S$
is a \textit{hitting set} of $\SC{CSTraces}(P)$. And the smallest fix is
a \textit{minimum hitting set} (MHS) of $\SC{CSTraces}(P)$. Note that MHS of any
collection of sets need not be unique.

\begin{definition} Given a set $U$ and a collection of sets $C \subseteq \mathcal{P}(U)
\setminus \{ \emptyset \}$, a set $H \subseteq U$ is a \textbf{hitting set} of $C$ if 
$\forall _{S_i \in C}\; S_i \cap H \neq \emptyset$. 
Furthermore, $H$ is called a \textbf{minimum hitting set} (MHS) if $C$ does not have a smaller hitting set.
\end{definition}

Finding an MHS is NP-complete, but for the problem instances that we generate, it is usually quite easy to
find an MHS.

\subsection{A First Approach \cite{yahav10popl}}
\label{Se:mhs}


\algref{mhs} describes an initial approach for finding the smallest set $S$ such that
$\tup{P,\textit{disable}(S)}$ is correct. This approach is inspired from the work of Vechev et al. \cite{yahav10popl}. 
Let $\verify$ be a verification tool. Given a program, $\verify(P)$ returns $\bug(t)$ if $P$ has a bug and the
error trace is $t$, else it returns $\correct$.

\begin{figure*}[t]
\begin{minipage}[t]{0.48\linewidth}
\begin{algorithm}[H]
\begin{scriptsize}
\caption {Minimum Hitting Set Solution}
\label{Alg:mhs}
\begin{algorithmic}[1]
\STATE \textbf{input:} Concurrent program $P$ instrumented with Boolean guards for \texttt{yield}s.
\STATE \textbf{output:}  Set $S$ of context switches, such that \inangleb{P,disable(S)} is correct.
\STATE $\phi := true $
\STATE $C := \emptyset$
\LOOP \label{algmhs:loopstart}
  \STATE res := $\verify$($\tup{P,\phi}$) 
  \IF{res == $\correct$} 
  \STATE \textbf{break}
  \ENDIF 
  \STATE \textbf{let} $\bug(t)$ = res
    \IF{$\CS(t) == \emptyset$} 
       \STATE \textbf{throw} exception(``Program has a sequential bug $t$'')
    \ENDIF
    \STATE $\phi := \phi \wedge \displaystyle \left( \bigvee _{cs \in CS(t)} \neg cs\right) $ \label{algmhs:phi}
    \STATE $C := C \cup \{ \CS(t) \}$ 

\ENDLOOP \label{algmhs:loopend}
\RETURN $\textit{MHS}(C)$
\end{algorithmic}
\end{scriptsize}
\end{algorithm}
\end{minipage}
\hspace{0.5cm}
\begin{minipage}[t]{0.48\linewidth}
\begin{algorithm}[H]
\begin{scriptsize}
\caption {Optimized Minimum Hitting Set Solution}
\label{Alg:optmhs}
\begin{algorithmic}[1]
\STATE \textbf{input:} Concurrent program $P$ instrumented with Boolean guards for \texttt{yield}s.
\STATE \textbf{output:}  Set $S$ of context switches, such that \inangleb{P,disable(S)} is correct.
\STATE $\phi := true $
\STATE $C := \emptyset$
\LOOP 
  \STATE res := $\verify$($\tup{P,\phi}$) 
  \IF{res == $\correct$} 
  \STATE \textbf{break}
  \ENDIF 
  \STATE \textbf{let} $\bug(t)$ = res
  \IF{$\CS(t) == \emptyset$} 
     \STATE \textbf{throw} exception(``Program has a sequential bug $t$'')
  \ENDIF
  \STATE $C := C \cup \{ \CS(t) \}$ 
  \STATE $\phi := \textit{disable}(\textit{MHS}(C))$ \label{opt:phi}
\ENDLOOP \label{algmhs:loopend}
\RETURN $\textit{MHS}(C)$
\end{algorithmic}
\end{scriptsize}
\end{algorithm}
\end{minipage}
\end{figure*}

\algref{mhs} iteratively (lines \ref{algmhs:loopstart}-\ref{algmhs:loopend}) finds an error trace $t$ and
stores the set of context switches taken by it in $C$. Then $\phi$ is modified to make sure
that the same trace $t$ does not manifest again (line \ref{algmhs:phi}) by disallowing at least one of the context
switches taken by $t$. This is repeated until no more bugs are found. 
\algref{mhs} returns a solution of the smallest size by computing the MHS of $C$.

\begin{wrapfigure}{L}{0.5\textwidth}
\begin{scriptsize}
\begin{tabular}{ll}
\begin{minipage}{0.2\textwidth}
\begin{verbatim}
x = 10; 
if(cs1) yield(); 
assert(x == 10);
\end{verbatim}
\end{minipage} &
\hspace{0.1cm}
\begin{minipage}{0.3\textwidth}
\begin{verbatim}
if(cs2) yield(); 
x = 5; 
if(cs3) yield(); 
tmp = 1; 
if(cs4) yield(); 
tmp = 1; 
... 
if(csN) yield(); 
tmp = 1;
\end{verbatim}
\end{minipage}
\end{tabular}
\end{scriptsize}
\caption{A code snippet.}
\label{Fi:Code1}
\end{wrapfigure}

As mentioned in \sectref{Overview}, this algorithm is not very efficient. Consider the code
snippet shown in \figref{Code1}. Suppose there are two threads executing this code. The first
thread executes the code on the left and the second thread executes the code on the right. The
program fails whenever the statement \texttt{x = 5} gets interleaved between statements \texttt{x = 10}
and the assertion.  Assignments to \texttt{tmp} are redundant but they introduce extra yield points.

When we run \algref{mhs} on this code, we can get error traces that first execute \texttt{x=10},
then context switch at \texttt{cs1}, then execute \texttt{x=5} and some part of the second thread,
context switch at $\texttt{cs}_i$ (for $3 \leq i \leq N$), and then fail the assertion. There
can be $N-1$ such traces. Thus, \algref{mhs} will potentially make $N-1$ calls to $\verify$.
While the fix is to disable just \texttt{cs1}, the number of verification calls made by this approach is proportional
to the size of the program.

\subsection{Our Approach}
\label{Se:optmhs}

\algref{optmhs} improves the previous algorithm by being more efficient when the size of the solution
is small. The main difference is on line \ref{opt:phi}. It computes a proposed solution by looking at
all previous traces. To see why this is an improvement, let us again consider the program in \figref{Code1}.
Suppose the first trace takes context switches $S_1 = \{\texttt{cs}_1, \texttt{cs}_5\}$. 
Then $C = \{ S_1 \}$ and it has two 
possible choices of MHSs. Suppose (unluckily) we pick $\textit{MHS}(C)$ as $\{\texttt{cs}_5\}$. Then $\phi$ disables
$\texttt{cs}_5$. The $\verify$ call will return another error trace passing through, say, 
$S_2 = \{\texttt{cs}_1, \texttt{cs}_8\}$ (note that
all error traces have to take $\texttt{cs}_1$). Then $C = \{ S_1, S_2 \}$ and has exactly one MHS, 
which is $\{\texttt{cs}_1\}$. 
Thus, we converge to the desired solution in just two queries, independent of $N$. Furthermore, the constraints
$\phi$ added to the program $P$ are much simpler than the ones added by \algref{mhs}, making the job of the
verifier easier.

\begin{theorem} Given a program $P$ with no sequential bugs, \algref{mhs} and
\algref{optmhs} compute a
minimum hitting set of \SC{CSTraces}($P$). 
\end{theorem}
\textit{Proof.} 
Let $m$ be the MHS of \SC{CSTraces}. Each of the algorithms returns an MHS over some subset of \SC{CSTraces}.
Let $C_i$ be the subset used by Alg. $i$ and let $m_i$ be its MHS. 
Both $m_1$ and $m_2$ are valid fixes because $\verify$ eventually returns $\correct$.
Thus, both are hitting sets of \SC{CSTraces}.
Because $m$ is an MHS of \SC{CSTraces}, it must be a hitting set of $C_1$. 
This implies $|m_1| \leq |m|$. Thus, $m_1$ is an MHS of $\SC{CSTraces}$.
Same argument applies for $m_2$.

\subsubsection{Performance comparision between \algref{mhs} and \algref{optmhs} : }
\label{Se:compare}

A direct theoretical comparison between the running times of \algref{mhs} and
\algref{optmhs} is
difficult because of inherent non-determinism in these algorithms. In particular, the
verification tool may return any
arbitrary buggy trace in the program fed to it, making it possible for any of \algref{mhs} and
\algref{optmhs} to get "lucky" and converge to a fix faster.
However, we can show that if both algorithms witness the same set of traces, then
\algref{optmhs}
is never worse than \algref{mhs}.

Let $\phi^{alg1}$ and $\phi^{alg2}$ denote the constraints generated by \algref{mhs} and
\algref{optmhs} respectively, on lines \ref{algmhs:phi} and \ref{opt:phi}. Further, suppose that the first $n$
iterations of the algorithms
witness the same traces $t_1,\dots,t_n$. Then it must be that $\phi^{alg2}$ is stronger
than $\phi^{alg1}$: For every trace $t_i$, $\phi^{alg1}$ has a clause $\inroundb{\bigvee
_{cs \in
CS(t_i)} \neg cs}$. On the other hand, $\phi^{alg2}$ has a clause with a single literal
$\neg cs'$, where, $cs'$ is the context switch taken by $t_i$ and is a part of an MHS
computed by it. Then $\phi^{alg2} \rightarrow \phi^{alg1}$ follows from $a \rightarrow
a\vee b$ (for each clause corresponding to a trace) as well as $a \rightarrow b \wedge c
\rightarrow d \Rightarrow
a\wedge b \rightarrow c\wedge d$ (conjunction of clauses from all the traces).
Consequently, if \algref{mhs} terminates in the $n+1^{st}$ iteration, then so will
\algref{optmhs}. Our
experiments(\sectref{Experiments}) show the superiority
of \algref{optmhs} in practice.

\section{Weak atomicity inference}
\label{Se:Weak}

Computing a fix using weak atomicity is harder because it doesn't directly allow us to 
disable context switches. We set up some terminology first.

\begin{definition}
Given a trace $t$ and a context switch \texttt{cs} taken by
$t$, let $T$ be the thread that was executing when \texttt{cs}
was taken. Then the \textbf{\emph{lifespan}} of \texttt{cs} in $t$ is defined
as the set of all instructions (or program points) on $t$ after
\texttt{cs} but before $T$ got control back. In other words, the
lifespan of a context switch is the contiguous sub-trace
between the two instructions of the same thread that surround the
context switch.
\end{definition}

For example, suppose $t$ = $[a_1; a_2; a_3; b_1; b_2; c_1; c_2; b_3; a_4]$, where $a_i$ denote
instructions of thread $1$, $b_i$ denote instructions of thread $2$, and $c_i$ denote instructions
of thread $3$. Then the lifespan of the context switch at $a_3$ is $\{ b_1, b_2, c_1, c_2, b_3 \}$.

\begin{wrapfigure}{L}{0.5\textwidth}
\begin{minipage}{0.5\textwidth}
\begin{algorithm}[H]
\caption{Conflict-Based Weak Atomicity Inference}
\label{Alg:conflictwa}
\begin{scriptsize}
\begin{algorithmic}[1]
\STATE  \textbf{input:} Concurrent Program $P$ instrumented with Boolean guards and $\lock$, 
and strong atomicity solution $S$
\STATE \textbf{output: } Weak atomicity solution.
\STATE $\phi$ := $disable(S)$
\STATE $C$ := $\{ \}$
\LOOP
\STATE res := $\verify(\tup{P, \phi})$
\IF {res == \correct}
\STATE \textbf{break}
\ENDIF
\STATE \textbf{let} $\bug(t)$ = res
\IF {$\WCS(t) == \emptyset$}
\STATE \textbf{throw} exception(``Program has a sequential bug $t$'')
\ENDIF
\STATE $C$ := $C \cup \{\WCS(t,S)\}$
\STATE $\phi$ := $disable(\textit{MHS}(C) \cup S)$
\ENDLOOP
\RETURN $S \cup \textit{MHS}(C)$
\end{algorithmic}
\end{scriptsize}
\end{algorithm}
\end{minipage}
\end{wrapfigure}

The way to rule out an error trace $t$ using weak atomicity is to pick two yield instruction 
$y_1$ and $y_2$ on the trace such that: $(1)$ $y_1$ appears before $y_2$; 
$(2)$ $t$ context switches at $y_1$ and $(3)$ the lifespan of the context switch at $y_1$ includes $y_2$.
In this case, we say that $y_2$ \textit{conflicts} with $y_1$.
Then including both $y_1$ and $y_2$ in a weak atomic block will render $t$ infeasible.
Moreover, this is the only way to disable a trace using weak atomic blocks (without inserting or deleting extra code).
In contrast, for strong atomicity, we only had to look at $y_1$. Thus, weak atomicity forces us
to identify the conflict between threads.

As before, we introduce a Boolean constant for every yield instruction in the program. Furthermore,
we introduce a global Boolean variable $\lock$ that is initialized to \textit{false}. If $cs$ is the
Boolean constant associated with a yield, then we transform it as follows:
\[ \texttt{\SSZ{yield;}} \Rightarrow \left\{ 
     \begin{array}{l}
       \texttt{\SSZ{if($\neg$cs) \{ assume lock == false; lock = true; \}}} \\
       \texttt{\SSZ{yield;}} \\
       \texttt{\SSZ{if($\neg$cs) \{ lock = false; \}}} 
     \end{array} \right.
\]
This way, setting a Boolean constant to \textit{false} is as if the corresponding \texttt{yield} is included
in a weak atomic block.

For a trace $t$, let $\WCS(t)$ be the set of $(cs_1, cs_2)$ pairs such that
$cs_i$ corresponds to a yield instruction $y_i$, and $y_2$ conflicts with $y_1$. Let $\SC{WCSTraces}(P)$
be $\{ \WCS(t) ~|~ t \in \SC{BTraces}(P) \}$. The smallest solution is given by the MHS of $\SC{WCSTraces}(P)$. If this
set is $W$, then the following set of yields need to be protected by a weak atomic block:
$\{ y ~|~ (y_1,y) \in W \text{ or } (y,y_2) \in W \}$.

We can now set up our algorithm in a similar fashion to \algref{optmhs}. However, we now have to gather 
pairs of instructions, which can lead to a large number of iterations. So we make use of a crucial optimization:
First, we compute the strong atomicity solution $S \subseteq \CSG$ for the program. 
Next, we only attempt to find the smallest extension of this solution that will fix the program 
using weak atomicity blocks. This is done as follows: For a trace $t$, instead of using $\WCS(t)$, we use
$\WCS(t,S) \eqdef \{ cs_2 ~|~ \exists cs_1 \in S: ~ (cs_1, cs_2) \in \WCS(t) \}$. Note that for an error
trace $t$, if $\WCS(t)$ is not empty then neither is $\WCS(t,S)$ because we know that some context switch
taken by $t$ belongs in $S$. Thus, we only look for conflicts with context switches in the strong
atomicity fix. Because $\WCS(t,S)$ is a subset of $\CSG$, we are again back to iterating over $\CSG$
rather than $\CSG \times \CSG$. \algref{conflictwa} formalizes this description.

The penalty of using this optimization is that we do not guarantee the smallest fix, however, we do guarantee
the smallest extension to the strong atomicity fix, and in our experiments we always obtained the smallest fix
possible. 

\section {Implementation and Experiments}
\label{Se:Experiments}

We have implemented \algrefs{optmhs}{conflictwa} in a tool called \tool. 
We use \SC{Poirot} \cite{poirot,corral} as the underlying verification tool. 
\SC{Poirot} is really a bug-finding tool; it searches over all behaviors
up to a bounded number of context switches, thus, it cannot prove the absence of bugs.
In this case, the fix returned by \tool is correct only up to the capabilities of 
\SC{Poirot}. In our experiments, we manually verified that the the computed fixes were sound.
In principle, we could have used a true verification tool like \SC{Threader} \cite{threader} 
inside \tool to obtain sound fixes.

\subsubsection{Results}
We evaluate the effect of changing various parameters on the performance of 
\algrefsp{mhs}{optmhs}{conflictwa}.
Consider the parameterized program shown in \figref{Code2}. It has two threads: the first executes the
code on the left and the second thread executes the code on the right. The program has three parameters
$p_1, p_2$, and $p_3$ that control the program size. Note that the strong atomicity fix is to enclose
the entire body of the first thread in an atomic block. Thus, the size of the strong atomicity fix is
$p_1 + 1$ (the number of yields inside this block of code). The size of weak atomicity fix is 
$p_1 + p_2$ because all of the assignments to \texttt{x} in the second thread must be put inside a weak atomic
block as well. The parameter $p_3$ controls the number of irrelevant assignments to shared variables. Results
are shown in \figref{param}. Here, \#CS is the number of yield instructions inserted in
the program, \#Q indicates the number of queries made to \SC{Poirot} and the last column
indicates running time in seconds. Compare \figref{psizeoptmhs} with \figref{psizemhs}. As
expected, \algref{mhs} requires more calls to \SC{Poirot} as the size of the
program increases. However, the number of calls made by \algref{optmhs} remains constant
irrespective of the program size. \figrefs{wsizeoptmhs}{ssizemhs} show that the number of
queries required by \algrefs{optmhs}{conflictwa} increases almost linearly as the size of the solution
increases. Here, (W) in columns indicates the numbers for weak atomicity fix.

\begin{figure}[t]
\centering
\begin{scriptsize}
\begin{tabular}{cp{.8cm}c}
\begin{minipage}{0.3\textwidth}
\texttt{x = 10;} \\
$\texttt{[y = 1;]}^{p_1}$ \\
\texttt{assert(x == 10);}
\end{minipage} & \large{$\displaystyle \parallel$} &
\begin{minipage}{0.3\textwidth}
$\texttt{[x = 1;]}^{p_2}$ \\
$\texttt{[y = 1;]}^{p_3}$ \\
\end{minipage}
\end{tabular}
\end{scriptsize}
\caption{A parameterized program with two shared variables: x and y. 
Here, $[\texttt{st}]^n$ denotes the statement \texttt{st} repeated $n$ times.}
\label{Fi:Code2}
\end{figure}

\Omit{
\begin{figure}
\begin{tabular}{ll}
\begin{minipage}{0.5\textwidth}
\texttt{x = 10;} \\
$\texttt{[y = 1;]}^{p_1}$ \\
\texttt{assert(x == 10);}
\end{minipage} &
\begin{minipage}{0.5\textwidth}
$\texttt{[x = 1;]}^{p_2}$ \\
$\texttt{[y = 1;]}^{p_3}$ \\
\end{minipage}
\end{tabular}
\caption{A parameterized program with two shared variables: x and y. 
Here, $[\texttt{st}]^n$ denotes the statement \texttt{st} repeated $n$ times.}
\label{Fi:Code2}
\end{figure}
}

\begin{figure}[t]
\begin{scriptsize}
\subfloat[Changing Program Size($p_3$) with \algref{optmhs} with $p_1=0,p_2=1$]
{
\Omit{
\begin{tabular}{|c|c|c|c|c|}
\hline \#CS &   \#Q & t(sec) & \#Q(W) & t(W)(sec) \\
\hline 
10 &  9 & 5.4 &  11 & 6.2 \\ \hline
30 &  9 & 7.8 &  11 & 9.0 \\ \hline
50 &  9 & 10.3 &  11 & 12.6 \\ \hline
70 &  9 & 12.9 & 11 & 15.7 \\ \hline
90 &  9 & 15.1 & 11 & 18.7 \\ \hline
\end{tabular}
}
\begin{tabular}{|c|c|c|c|c|c|}
\hline $p_3$ & \#CS &   \#Q & t(sec)  \\
\hline 
0 &  5 &  2 & 2.3  \\ \hline
10 & 15 &  2 & 2.6 \\ \hline
20 & 25 &  2 & 3.0 \\ \hline
30 & 35 &  2 & 3.2 \\ \hline
40 & 45 &  2 & 3.6 \\ \hline
\end{tabular}
\qquad
\label{Fi:psizeoptmhs}
}
 \qquad
\subfloat[Changing $p_2$, keeping $p_1=0,p_3=0$ \label{Fi:wsizeoptmhs} ]
{
\Omit{
\begin{tabular}{|c|c|c|c|c|c|}
\hline
WFIX & \#CS & \#Q & t(sec) & \#Q(W) & t(W)(sec) \\ \hline
4 & 63 & 10 & 7.58 & 15 & 12.1 \\ \hline
8 & 67 & 10 & 7.96 & 19 & 15.8 \\ \hline
12 & 71 & 10 & 8.28 & 23 & 20.8 \\ \hline
16 & 75 & 10 & 8.8 & 27 &  24.89 \\ \hline
20 & 79 & 10 & 9.38 & 31 & 29.44 \\ \hline

\end{tabular}
}
\begin{tabular}{|c|c|c|c|c|c|}
\hline
$p_2$ & \#CS & \#Q & t(sec) & \#Q(W) & t(W)(sec) \\ \hline
4  & 8 &  3 & 2.9 & 8 & 5.1 \\ \hline
8  & 12 & 3 & 3.0 & 12 & 8.5 \\ \hline
12 & 16 & 3 & 3.3 & 16 & 11.7 \\ \hline
16 & 20 & 3 & 3.5 & 20 &  15.8 \\ \hline
20 & 24 & 3 & 3.7 & 24 & 20.8 \\ \hline

\end{tabular}
} \qquad
\subfloat[Changing Program Size($p_3$) with \algref{mhs} with $p_1=0,p_2=1$ 
\label{Fi:psizemhs}
]
{
\begin{tabular}{|c|c|c|c|}
\hline $p_3$ & \#CS &   \#Q & t(sec) \\
\hline 
0 & 5 &  3 & 2.7 \\ \hline
10 & 15 &  13 & 7.7 \\ \hline
20 & 25 &  23 & 16.2 \\ \hline
30 & 35 &  33 & 27.0 \\ \hline
40 & 45 &  43 & 41.4 \\ \hline
\end{tabular}
\qquad
} \qquad 
\subfloat[Changing $p_1$, keeping $p_2=1,p_3=0$
\label{Fi:ssizemhs}
]
{
\Omit{
\begin{tabular}{|c|c|c|c|c|c|}
\hline

SFIX & \#CS & \#Q & t(sec) & \#Q(W) & t(W)(sec) \\ \hline
2 & 63 & 10 & 7.58 & 15 & 12.1 \\ \hline
4 & 65 & 21 & 7.58 & 26 & 36.74 \\ \hline
6 & 67 & 36 & 7.58 & 41 & 40.57 \\ \hline
8 & 69 & 55 & 7.58 & 60 & 52.1 \\ \hline
10 & 72 & 78 & 7.58 & 83 & 73.2 \\ \hline
\end{tabular}
}
\begin{tabular}{|c|c|c|c|c|c|}
\hline

$p_1$ & \#CS & \#Q & t(sec) & \#Q(W) & t(W)(sec) \\ \hline
0 & 5 & 2 & 2.3 & 4 & 2.9 \\ \hline
2 & 7 & 7 & 4.3 & 9 & 5.1 \\ \hline
4 & 9 & 11 & 5.9 & 13 & 6.9 \\ \hline
6 & 11 & 15 & 8.0 & 17 & 8.8 \\ \hline
8 & 13 & 19 & 10.1 & 21 & 11.2 \\ \hline
\end{tabular}
}
\caption{Effects of changing various parameters of the program in \figref{Code2}}
\label{Fi:param}
\end{scriptsize}
\end{figure}

Next, we ran \tool on various benchmarks gathered from previous work.
The results are shown in \tableref{res}. In the table, LOC is lines of code,
\#CS is the number yield instructions inserted in the program, Sol Size
is the number of program points as part of the computed fix, \#Queries is the number of
times \SC{Poirot} was called and the last column is the running time in seconds. The sub-columns
$S1$ and $S2$ indicates results for strong atomicity by \algref{mhs} and \algref{optmhs}
respectively.  $W$ indicates weak atomicity results obtained by running \algref{optmhs}
followed by \algref{conflictwa}. Numbers in bold indicates the better results amongst
\algref{mhs} and \algref{optmhs}. 
Against each benchmark, we refer to the paper from which it was
obtained. Here, \IT{banking\_inpaper} is the running example used in \figref{banking}.
Both the algorithms converged to the same solution for
strong atomicity for all the examples. On the average \algref{mhs} takes 20\% more
queries and 74\% more time as compared to \algref{optmhs}. If we discount for the outlier
benchmark "BankAccount", \algref{mhs} requires twice the number of queries on the average.
As mentioned in \sectref{compare} non-determinism plays a role as the two algorithms
witness different set of traces and takes different amount of time.
It is important to note that the most expensive operation in terms of time is a call to
$\verify$. We have observed that most of the time is spent inside the subroutine
$\verify$. Compared to this, the time consumed in computing MHS is negligible.
For all of the examples, we manually inspected  as well as cross verified with the papers
from which the benchmarks were taken. We found the quality of the solution proposed
by \tool to be the smallest and precise.
On the other hand, for programs \IT{logProcessNSweep} and \IT{CircularList}, 
the original approach \cite{recon2011pldi} proposes a fix that includes 
$1$ and $3$ extra program points, respectively, which are not relevant to the bug being fixed.

\begin{table}[t]
\centering
\begin{scriptsize}
\begin{tabular}{|l|r|r|r|r|r|r|r|r|r|r|}
\hline
\multirow{2}{*}{Example} & \multirow{2}{*}{LOC} & \multirow{2}{*}{\#CS} &
\multicolumn{2}{|c|}{Sol Size} & \multicolumn{3}{|c|}{\# Queries} & \multicolumn{3}{|c|}{Time(sec)} \\ \cline{4-11}
& & & S & W & S1 & S2 & W & S1 & S2 & W \\ \hline
banking\_inpaper(fig.\ref{Fi:banking}) & 62 & 22 & 2 & 3 & 22 & \MB{6} & 9 & 35.9 &
\MB{12.3} & 20.8 \\ \hline
banking\_inpaper\_corpus (fig. \ref{Fi:bankingcorpus}) & 58 & 23 & 3& 4 & 12 &  \MB{7} &9&
16.6 & \MB{11.4} &13.5 \\ \hline
apache1 \cite{thakur2009woda} & 64 & 10 & 2 &2& 4 &  \MB{3} & 4 & 4.5 & \MB{4.0} & 4.3\\ \hline
mozilla1 \cite{colorsafe} & 64 & 7 & 2 &2& 4 & \MB{3} &4& 3.7 & \MB{3.3} &3.6\\ \hline
mysql \cite{atomtracker} & 70 & 12 & 4 &6& 14 & \MB{13} & 16 & 9.4 & \MB{9.3} &10.8 \\ \hline
banking\cite{Wang2009fm} & 231 & 52 & 4 & 4 & 36 & \MB{23} & 24 &  298.5 & \MB{187.3} & 226.6 \\ \hline
defrag \cite{yahav10popl} & 142 & 37 & 2 & 2 & 56 &  \MB{5} & 6 & 475.5 & \MB{187.6} & 2384.1 \\ \hline
doubleLockQueue\cite{michael1996podc} & 144 & 29 & 4 & 4 & 8 &  \MB{7} & 8 & \MB{321.4} & 503.7 & 559.8 \\ \hline
jsClearMessagePane\cite{recon2011pldi} & 311 & 68 & 2 & 4 & 51 &  \MB{7} & 12 & 854.3 &
\MB{175.0} & 489.2 \\ \hline
jsInterpBufferBool\cite{recon2011pldi} & 215 & 36 & 1 & 2 & 23 &  \MB{3} & 15 & 85.6 &
\MB{16.3} & 65.9 \\ \hline
BankAccount\cite{recon2011pldi} & 149 & 32 & 12 & 14 & \MB{465} & 491 & 494 & 4306.0 &
\MB{2249.1} & 2546.5 \\ \hline
CircularList\cite{recon2011pldi} &  139 & 29 & 8 & 9 & \MB{9} &  \MB{9} & 11 & 232.8 & \MB{229.5} & 669.0 \\ \hline
StringBuffer\cite{recon2011pldi} & 126 & 25 & 11 & 12 & \MB{45} & 48 & 50 & \MB{202.4} & 205.0 &	524.8 \\ \hline
logProcessNSweep\cite{recon2011pldi} & 149 & 26 & 3 & 4 & 26 & \MB{23} & 26 & 371.3 &
\MB{332.8} & 474.5 \\ \hline
compute\cite{wypiwyg} & 51 & 7 & 2 & 2 & 6 &  \MB{5} & 6 & 9.7& \MB{8.8} & 14.9 \\ \hline
average\cite{wypiwyg} & 69 & 14 & 9 & 9 & \MB{24} & \MB{24} & 25 & \MB{16.3} & 16.4 & 17.4 \\ \hline
increment\cite{wypiwyg} & 27 & 5 & 1 & 1 & \MB{2} & \MB{2} & 3 & 3.2 & \MB{3.1} & 3.4 \\ \hline
nonDetRet\cite{wypiwyg} & 49 & 26 & 3 & 3 & 8 & \MB{6} & 7 & 13.2 & \MB{11.7} & 12.0 \\ \hline
\end{tabular}
\caption{Results of running \tool on a number of program snippets with published concurrency bugs.}
\label{Ta:res}
\end{scriptsize}
\end{table}

\Omit{
\section{Conclusions}
\label{Se:Summary}

This paper proposes techniques for automatically fixing concurrency bugs. The fix is reported in the form of
atomic block annotations, where the user can select either strong atomicity or weak atomicity. Atomic blocks
are usually simple and easy-to-understand description of synchronization. Our experiments show, on a collection
of benchmarks on concurrency bugs, that our technique is able to infer the ideal fix in each case.

\Omit
{While we infer atomic blocks that enclose the least number of program points, this is only a heuristic for allowing
maximum concurrency in the program and get the best possible  performance out of the fixed program. We cannot make 
guarantees about the performance. In fact, it is possible that our fixes may introduce deadlocks when the program
already has blocking synchronization. As future work, we would like to study techniques that not only fix a program
for safety properties, but also liveness properties.}

}

\bibliographystyle{plain}
\bibliography{biblio}
\end{document}